\documentstyle[epsf]{article}

\newcommand{\gsim}{\;\rlap{\lower 3.5 pt \hbox{$\mathchar \sim$}} \raise 1pt
 \hbox {$>$}\;}
\newcommand{\lsim}{\;\rlap{\lower 3.5 pt \hbox{$\mathchar \sim$}} \raise 1pt
 \hbox {$<$}\;}

\begin{document}

\title{
\vskip-3cm{\baselineskip14pt
\centerline{\normalsize\hfill MPI/PhT/97--80}
\centerline{\normalsize\hfill hep-ph/9711467}
\centerline{\normalsize\hfill November 1997}
}
\vskip1.5cm
${\cal O}(\alpha_s^2)$ Corrections to Current Correlators\thanks{
To appear in {\it Proceedings
of the International Europhysics Conference on High Energy Physics},
Jerusalem, Israel, 19-26 August 1997.
}
}
\author{Matthias\, Steinhauser
}
\date{}
\maketitle

\vspace{-1em}
\begin{center}
Max-Planck-Institut f\"ur Physik,
    Werner-Heisenberg-Institut,\\ D-80805 Munich, Germany
\end{center}
\vspace{.3cm}

\begin{abstract}
\noindent
In this contribution
three-loop QCD corrections to current correlators are considered.  
The application of the large momentum procedure, 
which provides a systematic expansion in $(m^2/q^2)^n$,
allows the computation of terms up to $n=6$ for the 
vector and axial-vector correlator and up to $n=4$ for the scalar
and pseudo-scalar correlator.
Some physical applications are discussed.
\end{abstract}

\vspace{.3cm}

There are plenty of important observables which are essentially
given by the transversal part of the current correlators, $\Pi(q^2)$,
with different tensorial structure.
The cross section for the annihilation of $e^+e^-$ to hadrons
is governed by the vector current correlator. The decay rate of the
$Z$ boson as well as the production of top quarks is a combination of
the vector and axial-vector correlator. 
For the decay rate of a scalar of pseudo-scalar Higgs boson
the correlators induced by scalar, respectively, pseudo-scalar currents
are important.

At one- and even at two-loop level results valid for arbitrary
external momentum, $q$, and quark mass, $m$, are available for
quite some time \cite{CKKRep}.
However, proceeding to three loops the situation is much more
difficult. In the high energy region until recently only terms up
to ${\cal O}(m^2/q^2)$ were available.
Our aim was to use a systematic approach which allows the computation 
of higher order mass corrections to the current correlators.
A well-defined algorithm is provided by the so-called large momentum
expansion (LMP)
\cite{Smi95}.
Whereas at one- and two-loop level it is still possible to apply
the LMP ``by hand'' for the three-loop calculation this is
very painful. In the vector case, e.g., from the 18 initial diagrams 
altogether 240 subgraphs, which have to be expanded in their small
quantities, are generated when the LMP is applied.
This is barely feasible by hand. Therefore we decided to automate the
LMP. Details can be found in
\cite{CheHarKueSte97}.

\begin{figure}[h]
\begin{center}
\begin{tabular}{ccc}
    \leavevmode
    \epsfxsize=5.0cm
    \epsffile[110 265 465 560]{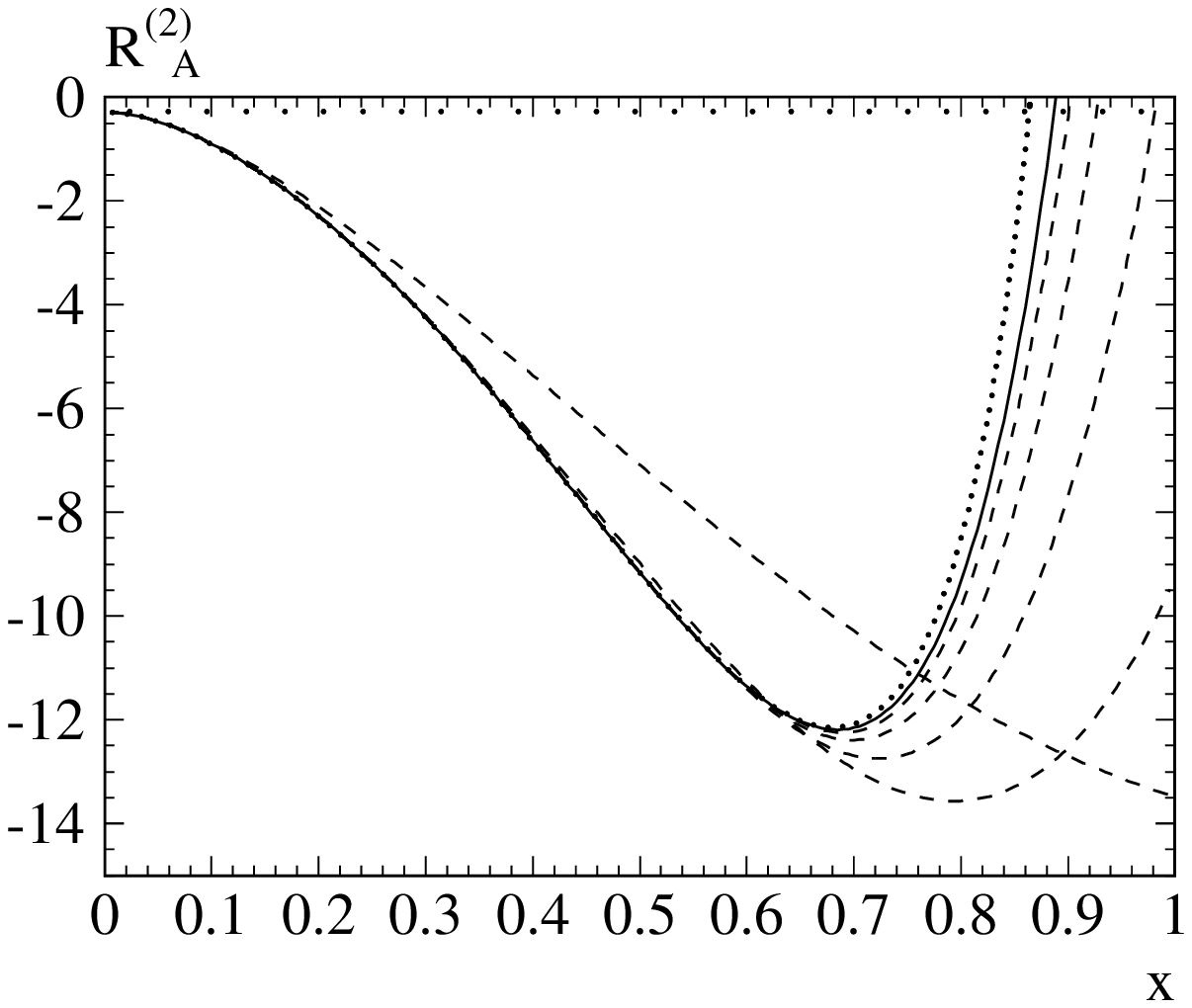}&
    \hphantom{xx}&
    \epsfxsize=5.0cm
    \epsffile[110 265 465 560]{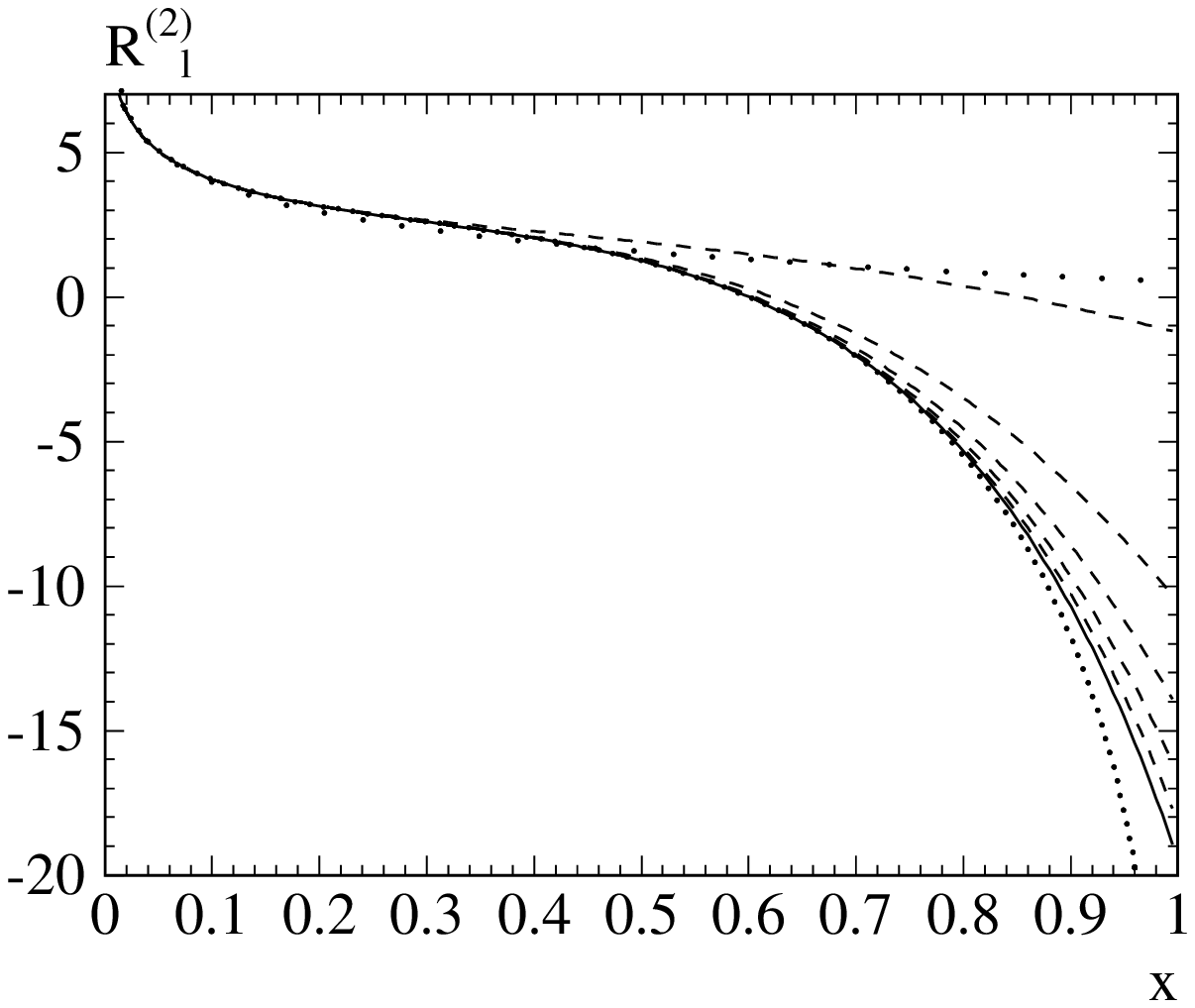}
\end{tabular}
    \caption[]{\label{figv}$R_A^{(2)}$ and $R_l^{(2)}$ plotted
      as a functions of $x = 2m/\sqrt{s}$.
      The scale $\mu^2 = m^2$ has been adopted.
      }
\end{center}
\end{figure}

The results of the ${\cal O}(\alpha_s^2)$ corrections to $\Pi(q^2)$ 
can be separated according to the different colour factors
and cast into the form:
\begin{eqnarray}
\Pi^{(2)} &=&
                C_F^2       \Pi_A^{(2)}
              + C_A C_F     \Pi_{\it NA}^{(2)}
              + C_F T   n_l \Pi_l^{(2)}
              + C_F T       \Pi_F^{(2)}
              + C_F T       \Pi_S^{(2)},
\nonumber
\end{eqnarray}
where $\Pi_S^{(2)}$ is the contribution from the singlet diagrams
which is not present in the vector case.
As an example in Fig.~\ref{figv} the results for the imaginary parts
of $\Pi_A^{(2)}$ and $\Pi_l^{(2)}$ in the case of
the vector correlator are shown. 
They contribute at ${\cal O}(\alpha_s^2)$ to the ratio
$R=\sigma(e^+e^-\to\mbox{hadrons})/\sigma(e^+e^-\to\mu^+\mu^-)$.
The wide-dotted line corresponds to
the massless corrections.
The dashed lines include successively higher orders in $m^2/s$ 
and finally the full curve comprises also the $(m^2/s)^6$ contributions.
The semi-analytical result~\cite{CKS} ($R_A^{(2)}$), respectively,
the exact result \cite{HKT} ($R_l^{(2)}$) is represented by
narrow dots.
The comparison shows that a very good approximation is obtained
up to $x=2m/\sqrt{s}\approx0.7$.

Similar results were obtained for the axial-vector corrections
\cite{HarSte972}. This allows the computation of the cross section
$\sigma(e^+e^-\to t\bar{t} +X)$ for energies $\sqrt{s}\gsim500$~GeV
(corresponding to $x\lsim0.7$ for $M_t=175$~GeV) up to 
${\cal O}(\alpha_s^2)$. For $\sqrt{s}=500$~GeV the NLO QCD corrections
amount to 2\%.

In \cite{HarSte97} the first five terms of the scalar and pseudo-scalar
corrections were evaluated.
This immediately leads to the inclusive decay rate of a scalar or
pseudo-scalar Higgs boson into top quarks. It was demonstrated that for a 
Higgs boson mass $M_H=450$~GeV the quartic corrections are still sizeable 
and of the same order of magnitude as the full ${\cal O}(\alpha_s^2)$
corrections. The contribution from the order $(m^2/s)^3$ and 
$(m^2/s)^4$ terms are small and confirm the validity of the expansion.

I would like to thank K.G. Chetyrkin, R. Harlander and J.H. K\"uhn
for the fruitful collaboration on this subject.

%

\end{document}